# A General Structural Order Parameter for the Amorphous Solidification of a Supercooled Liquid

Gang Sun[1,2]and Peter Harrowell[1,*]

1, School of Chemistry, University of Sydney, Sydney New South Wales 2006, Australia

2, Department of Fundamental Engineering, Institute of Industrial Science, University of Tokyo, 4-6-1 Komaba, Meguro-ku, Tokyo 153-8505, Japan.

[*] Corresponding author: peter.harrowell@sydney.edu.au

Abstract

The persistent problem posed by the glass transition is to develop a general atomic level description of an amorphous solidification. The answer proposed in this paper is to measure a configuration's capacity to restrain the motion of the constituent atoms. Here we show that the instantaneous normal modes can be used to define a measure of atomic restraint that accounts for the difference between fragile and strong liquids and the collective length scale of the supercooled liquid. These results represent a significant simplification of the description of amorphous solidification and provide a powerful systematic treatment of the influence of microscopic factors on the formation of an amorphous solid.

## 1. Introduction

For any solid to form, crystalline or amorphous, it must achieve a configuration in which each atom is highly restrained. This capacity of a configuration to mechanically restrain the displacements of the constituent particles represents an essential property of solid configurations. For particles interacting via short range repulsions, this restraint at zero temperature can be described by constraint counting [1-4]. A general structural measure for atomic restraint that can apply to any system of interaction particles at any temperature would represent an order parameter capable of describing any solidification and specifically, amorphous solidification. In this paper we shall pursue this line of reasoning.

To do so, we must address two questions. The first is the technical issue of defining such an order parameter. This is the subject of the first two Sections of the paper. We shall define a structural order parameter such that when it equals its maximum value (we can set this to one), the associated structure can be unequivocally identified as a low temperature solid. With this property, we can describe a structures approach to solidity based purely on the structural description alone, without having to explicitly demonstrate mechanical rigidity or high viscosity. While we would assume that the intermediate values of the structural order parameter correspond to some values of the stress relaxation time, our description of the change in the structure on cooling does not depend on establishing an explicit mapping, theoretically or empirically, between the structure and the dynamics.

Just to be quite clear, the amorphous solidification that we describe here is not the same as the glass transition. Regarding the latter transition, which is defined solely in terms of a change in dynamics, the goal is to explain the emergence of rigidity itself. In this paper, we follow the approach used in theories of freezing and consider the existence of the low temperature solid (amorphous, in this case) as a given, so that our account of solidification only needs achieve a relatively modest goal, to describe the appearance, on cooling, of those structural features that characterize the low temperature solid. The resulting decoupling of structure and dynamics is useful, we argue, allowing those aspects of amorphous solidification that can be expressed structurally to be studied unencumbered by complex relaxation kinetics.

The manner in which structure influences dynamics, remains a challenging problem. While we [5] and others [6] have established clear correlations between spatial heterogeneity of vibrational modes of amorphous groundstates and that of intermediate time dynamics, this correlation remains well short of a predictive description of the evolution of an amorphous configuration. The goal of this paper is to explore what progress we can make in describing the formation of glassy solids while avoiding the complexities of the associated kinetics.

If the description of amorphous solidification is not intended to provide an explanation of the timescale of relaxation dynamics, what *is* it good for? This is the second (and larger) question we shall address. To do this we must identify those features of the transformation of a liquid into a glass that *can* be recast into questions about the underlying structural change. An example of this situation is the fragility of a liquid. While conventionally associated with the temperature dependence of viscosity, fragility, as first conceived by Angell in 1985 [7], was introduced to describe a feature of the underlying structure of the liquid. In Section 5 of this paper, we shall confirm Angell's original expectation, i.e. that a liquid with a temperature dependent activation energy (a 'fragile' liquid) also exhibits a large temperature dependence in structure while another liquid with temperature -independent activation energy (a 'strong' liquid) exhibits a structure with only a weak dependence on temperature. A second example of dynamics being used to illuminate an underlying structural correlation are kinetic correlation lengths. In Section 6 we shall establish how correlation lengths arise directly as a consequence of the structural change associated with cooling the amorphous material.

Order parameters play a fundamental role in the statistical description of the transformation between two phases. While these transformations are often associated with thermodynamic singularities, an order parameter description is of greater generality and can be applied equally to solidification due to kinetic arrest without any associated singularity, unique transition point or even equilibrium. In cases where the phase transformation involves a change in symmetry of the microscopic structure, the choice of order parameter is usually dictated to be a measure of the structure associated with the broken symmetry. Magnetization in the Ising model or, in the case of the isotropic-nematic transition, the second moment of the distribution of orientation cosines, are examples of broken symmetry order parameters. In the case of more complex transformations, there may be multiple options for the order parameter. The freezing of a liquid into a crystal is such a case. In the 1979 density functional theory of Ramakrishnan and Yussouff [8], the order parameters are Fourier amplitudes of the singlet density. In 1984, Tarazona [9] proposed that the singlet density be treated as a sum of Gaussians centered on the crystal lattice sites so that the width of the Gaussian served as the order parameter for crystallization. While useful in the functional theories, these order parameters proved to be of little use in monitoring crystal-like fluctuations in the liquid. To address this problem, Steinhardt et al [10] introduced a bond order parameter based on the amplitudes of spherical harmonics of the local coordination geometry. This order parameter's evolution has continued with important modifications by Lechner and Dellago [11]. We present this history to establish the point that the choice of an order parameter, while constrained by the twin conditions that i) it quantifies a property defined for individual configurations, and ii) it clearly differentiates the two states involved in the transformation, retains a considerable freedom of choice. Any choice of order parameter is ultimately assessed and improved upon on the basis of its usefulness rather than some *ab initio* proof of worth, a point we ask the reader to keep in mind as we develop the order parameter in this paper.

Topological measures that generalize the bond order parameter of refs. [10,11] have been used to describe amorphous structures. These measures include Voronoi and Delaunay tesselation [12], common neighbor analysis [13] and tetrahedrality [14]. A significant number of these papers are not directly interested in the formation of the amorphous solid but, instead, are looking for the growth of non-crystalline structures that could kinetically obstruct crystal nucleation [15]. The role of local structure on dynamical arrest has been reviewed by Royall and Williams [16]. Perturbations that result in structural changes in a liquid have been found to also perturb relaxation kinetics [17,18]. While correlation between structure and dynamics is not disputed, it remains unclear whether a workable solid order parameter can be extracted from the local topology of a liquid. One difficulty is the huge diversity of local structure in a typical glass. Where crystals rarely exceed 4 distinct local coordination geometries [19], an amorphous solid can have over 100 local topological arrangements [20] that contribute significantly to the structure.

With a large multiplicity of local structures capable of contributing to restraint, we turn to restraint itself as providing the most generic criteria for quantifying an amorphous configuration. Tarazona's [9] Gaussian expansion of the density provides perhaps the first explicit measure of solidity in terms of an order parameter based on the degree of atomic restraint. His idea was extended to amorphous solids by Singh et al [21] who retained the Gaussian sum but replaced the crystal lattice with an amorphous structure. The thermodynamic treatment of amorphous solids, initiated in ref. [21], has evolved significantly over the years [22,23]. The associated order parameter is now defined [23] as the average of the overlap Q between two liquid configurations, $\vec{r}_\alpha^N$ and $\vec{r}_\beta^N$,

$$Q[\vec{r}_\alpha^N, \vec{r}_\beta^N] = \frac{1}{N} \sum_{i,j}^N w\left(\left|\vec{r}_{\alpha,i} - \vec{r}_{\beta,j}\right| / a\right) \qquad (1)$$

where w(y) is a step function, equal to 1 for y < 1 and 0 otherwise and $a$ is the resolving length. The average of Q could represent the average restraint about a reference configuration, with Q ~ 1 indicating that particles in the reference state are well restrained, at least under the sampling used to calculate the average. The problem is that, with the exception of the crystal structure, most choices of reference state and sampling ensemble result in <Q> = 0. Franz and Parisi [24] proposed a modified sampling in which they imposed an artificial potential V(Q) defined as the minimal work needed to keep a system at temperature T at a fixed value of Q from a typical equilibrium configuration of the same system. On lowering T, a minimum in V at a $Q^*\neq 0$ appears [25] representing the existence of an underlying metastable state of the unperturbed Hamiltonian, one characterized by a mean restraint $Q^*$ about the reference configuration. This formalism has recently been extended to address a localized measure of overlap were the Franz-Parisi potential is only applied to particles inside a selected spherical sub-volume [26]. While the formalism of ref. [24-26] represents an elegant strategy for the thermodynamic treatment of metastable states, it is not appropriate for the task of mapping a restraint order parameter in an instantaneous configuration.

## 2 The Configurational Restraint Parameter

We define the degree of restraint experienced by particles in an instantaneous configuration in terms of the equilibrium amplitude of the individual particle displacements $<\Delta r_j^2>_{eq}$ in a harmonic approximation of the Hamiltonian about that configuration. (Note that we have defined here the meaning of the term 'restraint'. This definition is does not depend on whether our property actually reproduces the dynamical effect of a particle trapped by a persistent cage of its neighbors. We shall return to this distinction later in the paper.)  Other measures are possible but this one is efficient to calculate and easily generalizable to any differentiable potential. To generate the harmonic approximation, we shall calculate the instantaneous normal modes (INM's) and their associated eigenvalues. The INM analysis alone, however, will typically not generate the harmonic potential we are after due to the presence of negative eigenvalues. These modes, which we shall refer to as 'unstable', have been the central focus of much of the literature on the harmonic analysis of liquids due to their possible role in particle dynamics [27,28]. In this paper we employ the INM's for a quite different purpose. Instead of treating the INM's as representations of the actual short time dynamics of the liquid, we use them as a tool for resolving the capacity of a configuration to restrain motion and, hence, generate our order parameter. From this perspective, each unstable mode simply represents an absence of information concerning the bounds on the fluctuations of the associated mode amplitude, a lack that we shall address using anharmonic contributions as explained below.

Previously, characterizations of the glass transition have been developed based on an estimation of the local Debye-Waller (DW) factor, $< \Delta r_j^2 >_\tau$, the mean squared displacement for some fixed short time interval τ. In contrast, $< \Delta r_j^2 >_{eq}$, calculated from the harmonic Hamiltonian (described below) is not a dynamic quantity. It is obtained by analysis of a single instantaneous configuration only and, therefore, can be regarded as a structural measure. As we shall demonstrate, $< \Delta r_j^2 >_\tau \neq < \Delta r_j^2 >_{eq}$ except at very low T. A number of proposals use the local Debye-Waller (DW) factor, not to describe the structure, but to estimate, either directly or implicitly, the temperature dependence of a relaxation time [29-32]. The local DW factor has been used explicitly as an order parameter in some recent studies. Yang et al [33] made use the local DW factor calculated from the dynamical matrix. Ding et al [34] introduced a composite order parameter consisting of the product of the Voronoi volume and the local DW factor. Schoenholz et al [35] have constructed a structure-based parameter optimized to differentiate a set of particles identified as being on the verge of relaxing from a set in persistent stable arrangements.  Finally, Tong and Xu [36] and Smessaert and Rottler [37] quantified the structural heterogeneity of the glass structures using a measure proportional to $<\Delta r_j^2>_{eq}$, calculated from the normal mode analysis, but only for the T = 0 glass structure. While acknowledging the similarities of ref. [33-37] with the present work, we point out two differences that will prove significant. The first is methodological – our use of $<\Delta r_j^2>_{eq}$ rather than $<\Delta r_j^2>_{\tau}$ for all temperatures allows us to completely remove dynamics from the definition of the order parameter and, as we shall show, provides a more complete characterization of structure. The second difference is conceptual.  These previous papers have focused almost solely on establishing a correlation between structure and dynamics.  We argue, instead, that the utility of an order parameter is established by its capacity to differentiate the high and low temperature structures as simply as possible.

As we do not make any use of dynamic information, we can sample configurations using Monte Carlo (MC) algorithms. Unless otherwise indicated, all calculations have been carried out using the $A_{80}B_{20}$ Lennard-Jones mixture introduced by Kob and Andersen [38] with 5000 particles in a simulation cell with periodic boundary conditions. Configurations are generated using particle swap MC simulations [39] at constant NVT with a fixed number density of 1.2. From the point at which the average potential energy deviates from linear dependence on T (see Supplementary Material) we determine that the liquid is equilibrated down to T = 0.36, below which the configurations we generate are no longer at equilibrium.

In Fig. 1a we plot the temperature dependence of the fraction $f_u$ of unstable modes (i.e. those with negative eigenvalues). We find that the fraction of unstable modes decreases monotonically with decreasing T and vanishes at T ~ 0. As shown, our values of $f_u$ are in good agreement with the previous calculations of Donati et al [40]. We calculate $< \Delta r_j^{\,2} >_{eq}$ in terms of the amplitudes $A_\alpha$ of the instantaneous normal modes via

$$< \Delta r_j^{\,2} >_{eq} = \sum_\alpha < A_\alpha^{\,2} >_{eq} (\vec{v}_\alpha^{\,j})^2 \qquad (2)$$

where $\vec{v}_\alpha^{\,j}$ is the components of the $\alpha$ eigenmode arising from particle j.

The mean squared mode amplitude $< A_\alpha^{\,2} >_{eq}$ can be evaluated by numerically integrating

$$< A_\alpha^{\,2} >_{eq} = \frac{\int A_k^{\,2} e^{-\beta U(A_\alpha)} dA_\alpha}{\int e^{-\beta U(A_\alpha)} dA_\alpha} \qquad (4)$$

where $U(A_\alpha)$ is the potential energy calculated as a function of the variation of the amplitude $A_\alpha$, holding all other mode amplitudes at zero and $\beta = (k_B T)^{-1}$. An example of the potential profile $U(A_\alpha)$ for an unstable mode is presented in Fig. 1b where we note that, with the inclusion of the anharmonic contributions, $U(A_\alpha)$ provides a well-defined bound on the mode amplitude fluctuations, even when the harmonic term does not. A useful shortcut would be to replace the numerical integration with the analytic harmonic expression

$$< A_\alpha^2 >_{eq} = k_B T / \lambda_\alpha \tag{3}$$

for modes with a positive eigenvalue $\lambda_\alpha > 0$. We shall test the accuracy of this analytic result below. The evaluation of $< A_\alpha{}^2 >_{eq}$ allows us to construct a local harmonic basin about any instantaneous configuration by replacing each negative eigenvalue by an effective positive one defined as $\lambda_{\alpha,eff} = (\beta < A_\alpha^2 >_{eq})^{-1}$, leaving us with a harmonic Hamiltonian,

$$H_{LH} = \frac{1}{2} \sum_\alpha^{stable} \lambda_\alpha A_\alpha^2 + \frac{1}{2} \sum_\alpha^{unstable} \lambda_{\alpha,eff} A_\alpha^2 \tag{5}$$

$H_{LH}$, we propose, represents as complete an account of the characteristics of a given configuration as can be extracted from that configuration only.

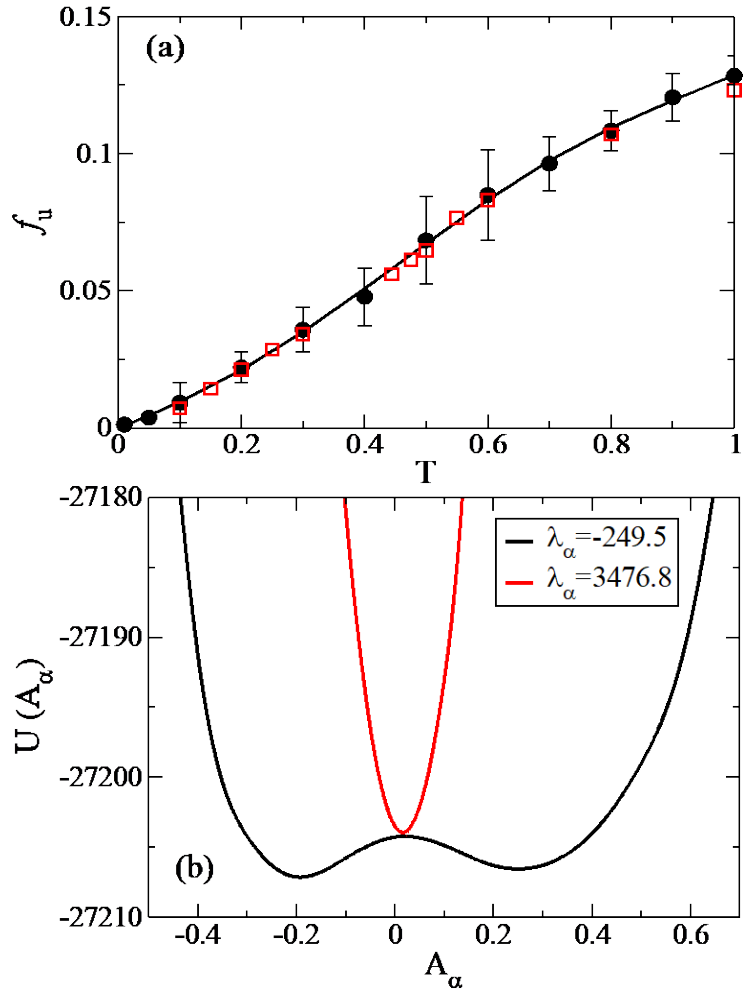

**Figure 1.** a) The fraction $f_u$ of unstable modes as a function of T for the KA liquid. The black circles are from the present paper while the red squares are data from ref. [40]. The vertical lines represent one standard deviation above and below the mean. b) Examples of the potential energy surface U(A) associated with the variation in amplitude $A_\alpha$ of an unstable (black line) and stable (red line) mode.

While the quantity $< \Delta r_i^2 >_{eq}$, as calculated using Eq. 2, provides a local measure of the configurational restraint implicit in the harmonic Hamiltonian, it is not the optimal choice for an order parameter. It is unbounded and, once $< \Delta r_i^2 >_{eq}$ exceeds the atomic length (squared), its value is of little significance since all such cases will be designated 'unrestrained'. For these reasons, we define, instead, an order parameter $\mu_j$ for the restraint of atom j as

$$\mu_j = \exp\left(-\frac{q^2 < \Delta r_j^2 >_{eq}}{6}\right) \tag{6}$$

where the average $<\ldots>_{eq}$ is taken over the local harmonic Hamiltonian and q is the magnitude of the wavevector of the first peak in the structure factor S(q), a value that varies little with temperature under the constant volume constraint (see Supplementary Material). By construction, the values of restraint lies in the interval $0 \le \mu_i \le 1$. In Fig. 2a we plot the average restraint $< \mu > = \frac{1}{N} \sum_i \mu_i$ as a function of T. In comparing the values of <μ> calculated with and without the analytic short cut for the stable modes in Eq. 3, we find no significant difference. In all subsequent calculations of μ we shall use Eq. 3 for the stable modes.

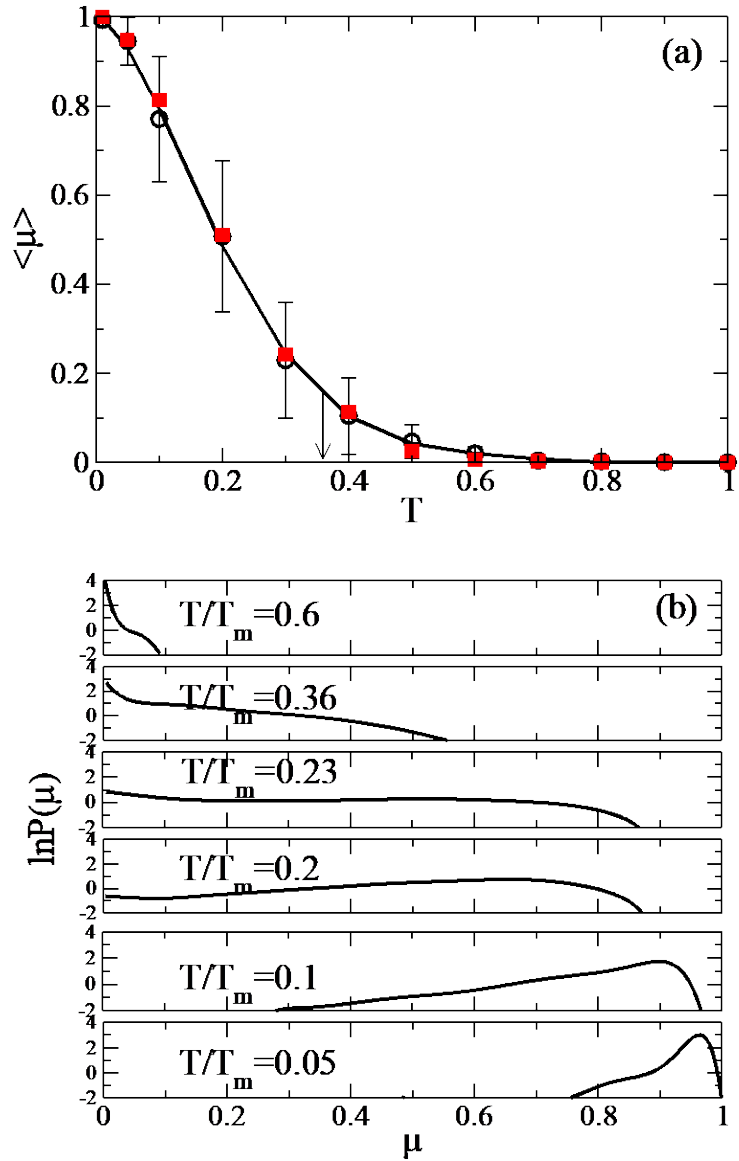

**Figure 2.** a) The plot of the average restraint <μ> as a function of T for the KA liquid. Results are presented based on purely the numerical integration of Eq. 2 (filled symbol) and using Eq. 3 for the stable modes (open symbol). The temperature, T = 0.36 where liquids fall out of equilibrium is indicated by arrow. The vertical lines represent one standard deviation above and below the mean. b) The distribution of μ for a range of temperatures, as indicated. For reference, the freezing point of the KA mixture is $T_m = 1.02$ [42].

We find that the average restraint exhibits a nonlinear increase with decreasing T. We emphasize that the temperature dependent changes in the order parameter μ describe a purely structural change in the instantaneous configurations during the transition from liquid to amorphous solid, so that the increase in <μ> on cooling represents an explicit measure of the progress towards the low temperature solid state characterized by <μ> =1.

The evolution of the structure into that of the amorphous solid is accompanied by the appearance of a broad binodal distribution in μ with peaks at 0 and ~ 0.6 as shown in Fig. 2b. The variance of the distribution exhibits a maximum at temperature T ~ 0.23, providing us with a structure-based characteristic temperature for the passage from liquid to solid, one associated with the maximum in the structural heterogeneity. We have established that the simulations fall out of equilibrium at T ~ 0.36 (see Supplementary Material). This latter temperature is determined solely by the relative times scales of relaxation and data sampling. The observation that the structural change occurs at a lower temperature than that associated with falling out of equilibrium is consistent with the idea that, while increased structural restraint is expected to contribute to slowing down, it is not necessary, a point we shall return to in the context of strong liquids.

## 3. The Anisotropic Accumulation of Restraint

The restraint order parameter can be extended to resolve the local anisotropy of the restraint imposed by a configuration. It has been noted previously [43] that a dynamic heterogeneity can take the form of an object with a lower dimension that that of the space, suggesting that the associated structural restraints are strongly anisotropic. The anisotropy of restraint is clearly of importance when heterogeneities such as interfaces [44] are present. Generalizing Eq.2, we can write

$$< x_i y_i > = \sum_\alpha < A_\alpha^2 > v_{\alpha,x}^i v_{\alpha,y}^i \qquad (7)$$

where $v_{\alpha,x}^i$ is the x-component of the contribution of particle i to the αth eigenvector. Diagonalizing the resulting variance matrix corresponds to carrying out a principal component analysis where the three eigenvalues - $\Gamma_1$, $\Gamma_2$ and $\Gamma_3$ – correspond, respectively, to the mean squared amplitude along the principal direction of motion, the next major direction orthogonal to this principal axes and the remaining direction orthogonal to the first two. The equilibrium mean squared displacement $< \Delta r_i^2 >_{eq} = \Gamma_{i,1} + \Gamma_{i,2} + \Gamma_{i,3}$ and the associated order parameters are defined as $\mu_{i,k} = \exp(-q^2 \Gamma_{i,k} / 6)$ so that

$$\mu_i = \mu_{i,1} \times \mu_{i,2} \times \mu_{i,3} \qquad (8)$$

We can now consider the anisotropy resolved description of amorphous solidification where we follow the temperature dependence of each component of restraint, as plotted in Fig. 3. The local anisotropy of the restraint on atom i can be quantified by the principal local anisotropy $\Lambda_{i,1} = (\Gamma_{i,1} - \Gamma_{i,3}) / \Gamma_{i,1}$ and the minor anisotropy $\Lambda_{i,2} = (\Gamma_{i,2} - \Gamma_{i,3}) / \Gamma_{i,2}$ whose variation with temperature is presented in Fig.3b.

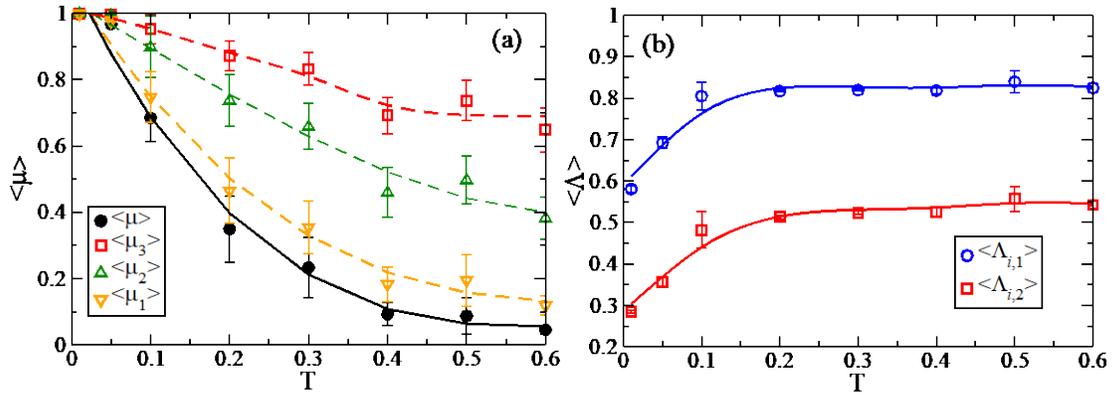

**Figure 3**. a) A plot of the temperature dependence of $\langle\mu_1\rangle$, $\langle\mu_2\rangle$ and $\langle\mu_3\rangle$ as defined in the text (dashed curves). Also included is the total average, $\langle\mu\rangle$ (solid curve). b) A plot of the principal and minor local anisotropies, $\Lambda_{i,1} = (\Gamma_{i,1} - \Gamma_{i,3})/\Gamma_{i,1}$ and $\Lambda_{i,2} = (\Gamma_{i,2} - \Gamma_{i,3})/\Gamma_{i,2}$ respectively, vs T. Lines are included as a guide to the eye.

As shown in Fig.3a, restraint in the weakly supercooled liquid (i.e. T > 0.4) is very anisotropic. The majority of particles are partially restrained and fluidity coexists with this high degree of low dimensional restraint. The dominance of the least restrained degree of freedom on the overall value of μ is a consequence of the multiplicative combination,

$$\mu = \prod_{i}^{3} \mu_i$$ . As shown in Fig.3b, this anisotropy is essentially independent of T down to

temperatures well with the amorphous solidification process (i.e. the range over which $\langle\mu\rangle$ exhibits a large increase on cooling in Fig. 2). The transformation between liquid and amorphous solid, therefore, is largely about the temperature dependence of restraint of the local soft direction for each particle. As $\mu_2$ also exhibits a significant, if smaller, temperature dependence, our results support the idea that amorphous solidification can be regarded as a 'stiffening' of a particle displacements within a sb-volume of dimension between one and two.

## 4. Configurational Restraint and the Self Intermediate Scattering Function

In supercooled liquids, the self-intermediate scattering function $F_s(q,t)$,

$$F_s(q,t) = \frac{1}{N}\left\langle \sum_{j}^{N} \exp\left(i\vec{q}\cdot\Delta\vec{r}_j(t)\right)\right\rangle \tag{9}$$

taken at a value of the wavevector q corresponding to the first peak of the structure factor S(q), is referred to as the structural relaxation function. The decay of this function in the supercooled liquid, as shown in Fig. 4a, proceeds via a two step process; a fast relaxation to a plateau value, $h$, which is subsequently relaxed by slower mechanism, the α process, with a strong temperature dependence. Previously, in the plateau height h has been identified as a 'non-ergodicity' parameter and linked to an order parameter developed for spin glasses [45]. In the limit of a harmonic solid, it can be shown that $h =< \mu_i >$ which raises the question, whether <μ> is essentially equivalent to the plateau value $h$ of $F_s(q,t)$ for all T?

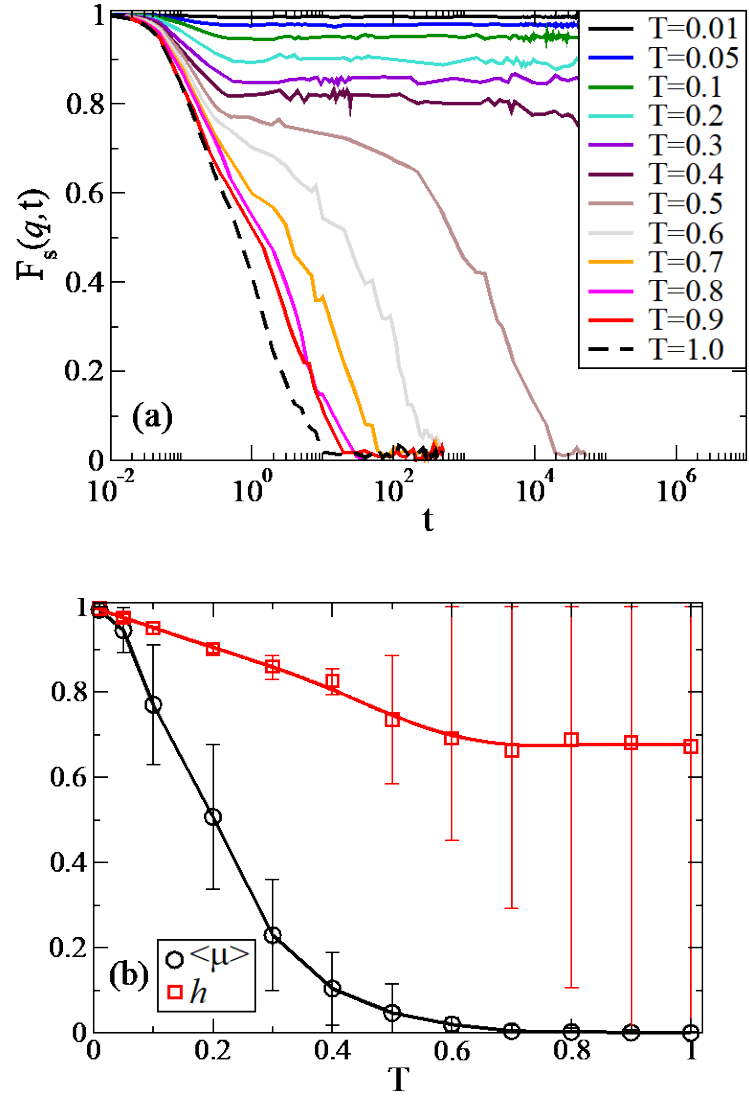

**Figure 4.** a) Plots of the self intermediate scattering function $F_s(q,t)$ vs log time for range of temperatures for the KA liquid. b) Plot of the plateau height $h$ of $F_s(q,t)$ and $\mu$ as a function of T. Note the large uncertainty in $h$ at high T due to the lack of a well-defined plateau. Details of the calculation of $h$ and the associated uncertainty are provided in the Supplementary Materials.

In Fig. 4b we establish that $< \mu > \neq h$, with the restraint <μ> exhibiting a much stronger dependence on temperature than $h$, dropping to zero over a temperature range for which the plateau exhibits only a weak variation. We also note that $h$ becomes ill-defined for T > 0.6 due to the disappearance of the plateau. The fact that $< \mu >$ is smaller than $h$ is perhaps surprising since the former corresponds to an average over a highly restricted set of fluctuations while the latter has no imposed constraints. The explanation is that the large amplitude fluctuations responsible for the small values of μ are not accessible to the dynamics on the lifetime of given instantaneous configuration. The short time dynamics cannot 'see' the softest modes supported by the individual configurations due to this motional 'stiffening'. This observation underlines the point, already made, that $< \mu >$ is not presented as a predictor of short time dynamics but as an order parameter whose essential function is to differentiate fluid and solid amorphous structures.

## 5. On the Difference in Configurational Restraint between Strong and Fragile Liquids.

As discussed in the Introduction, having defined a sensible order parameter we still have to establish just what phenomena can be usefully treated from this structural perspective. Our first candidate will be the striking difference between the glass transitions of oxide glass formers such as $SiO_2$ and the transition in liquids governed by largely isotropic interactions (i.e. molecular and metallic liquids). A measure of this difference is the temperature dependence of the effective activation energy for relaxation. In the case of the network-forming liquids ('strong' in Angell's classification) the activation energy is independent of T while in molecular liquids ('fragile'), the activation energy increases nonlinearly with decreasing temperature [7]. Can the description of these two transitions based on the order parameter μ capture an underlying structural difference?

Molten silica has been modelled with molecular dynamics using the SHIK-1 potential [46] which includes long range Coulomb interaction. The use of the PPPM method [47], a Fourier-based Ewald summation method, to calculate the Coulomb contribution to the potential energy, complicates the analytic calculation of the Hessian. We have used numerical differentiation to calculate the second derivations of the energy (details provided in the Supplementary Information). The fraction of imaginary modes in the model silica was found to be small, < 0.08 over the T range studied. The treatment of these modes and the subsequent calculation of the restraint $\{\mu_i\}$ was carried out as for the Lennard-Jones system. To usefully compare our two model liquids, we have plotted the results in Fig. 3 in terms of a reduced temperature $T/T_m$, using the following values of the melting points: for SHIK-1 silica $T_m =$ 2400K [48] and for the KA alloy, $T_m = 1.02$ [42]. These results have appeared in a preliminary report [41] and we include them here to provide the reader with a complete and coherent account of this new structural approach.

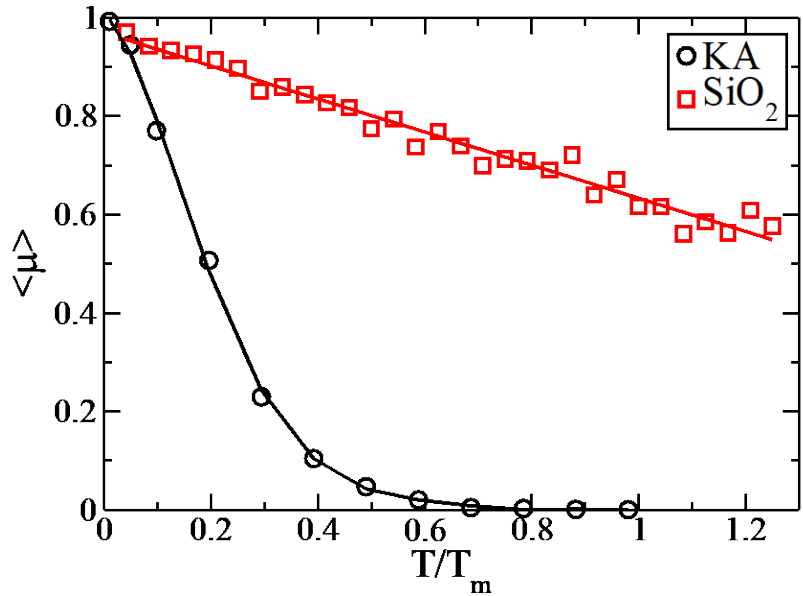

**Figure 5.** Plot of <μ> vs T/T$_m$ for the KA mixture and for silica. The temperature has been scaled by the respective melting points T$_m$ (provided in the text) solely to allow comparison between the data from the different liquids.

The difference in the temperature dependence of the configurational restraint between silica and the atomic mixture is striking. Despite the difference in the values of N, both systems are, close to their large N limiting behavior, based on the respective correlations lengths obtained previously [41] (see Supplementary Material), so that the difference represents an intrinsic material property. Where the atomic mixture exhibits an abrupt increase in restraint at a temperature well below the melting point, silica maintains a high degree of restraint all the way up to the boiling point. This result, suggestive as it is, does not explicitly allow us to account for the difference in activation energies. It does, however, allow us to identify a *structural* fragility that appears to align with the conventional *dynamic* fragility. Angell has proposed an overarching account of the glass transition based on the existence of just such a structural fragility in the form of an underlying continuous order-disorder transition [49]. In this picture the distinction between fragile and strong liquids is determined by whether this transition occurs below or well above $T_m$, respectively. There are obvious parallels between our results and Angell's proposal. The differences are that a) we present an explicit measure of the structural change associated with the underlying transformation and b) our transformation is not a cooperative thermodynamic order-disorder transition, a result we shall establish in Section 5.

What we demonstrate here is that these liquids, the atomic alloy and silica, are differentiated by their structural fragility. The structure of a strong liquid, even at equilibrium, lies close enough to that of the amorphous solid that there is little room for further development on cooling, the very development that characterizes the fragile liquid. One consequence of this result is the observation that, in the case of silica, the liquid is, structurally, indistinguishable from that of a hot solid. The inability of our order parameter to identify the liquid state in this case is, perhaps, not surprising. The viscosity of silica at its melting point is $\sim 10^5$ times the corresponding values for most fragile liquids. In the context of current timescales accessible to MD simulation, the silica melt does indeed resemble a solid at $T_m$.

## 6. Amorphous Solidification under Imposed Restraint: Pinning and Finite Size Effects

As we have defined it, the structural transformation from liquid to amorphous solid depends, not on the appearance of special topologies but, rather, on the presence of *any* structure capable of exerting local restraint. We can adjust this restraining influence, without change to the configurational geometry, by pinning random particles. Let $M = (1-c)N$ be the number of unpinned particles where c is the fraction of pinned particles. Following ref. 50, the impact of pinning on the instantaneous normal modes is calculated by diagonalizing the $3M \times 3M$ Hessian matrix consisting of only derivatives of coordinates of the unpinned particles. Note that the pinning is only applied when calculating the normal modes - not during the MC simulation used to generate the configurations.

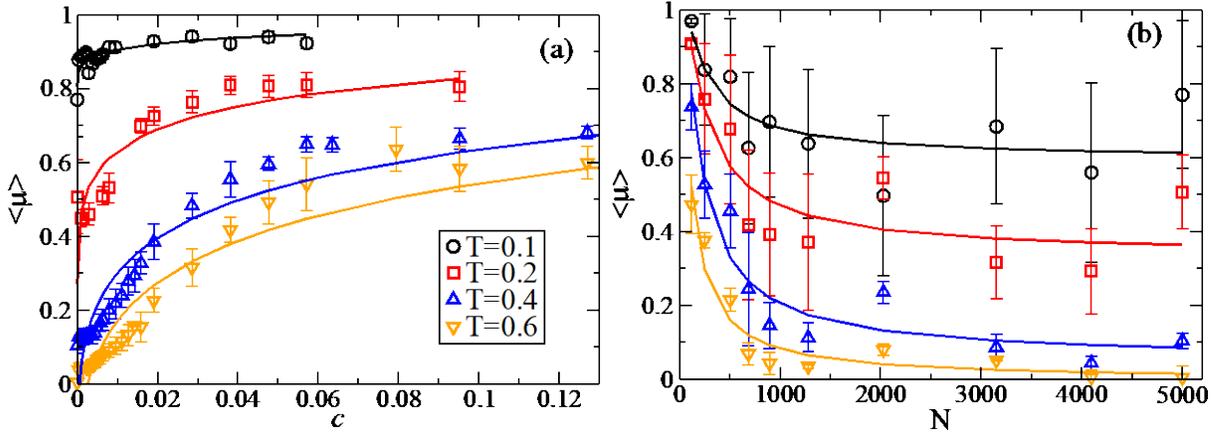

**Figure 6.** The dependence of the average restraint <μ> as a function of a) the fraction of c pinned particles and b) the system size N, for a number of temperatures for the KA liquid as indicated. The solid curves correspond to the curve in Eq.10 and 11, respectively, with fitted lengths ξ and ζ listed in Table 1. (The results in Fig.6a differ from those published previously [41] due to the unintended inclusion of center of mass motion in pinned systems in the earlier paper.)

As shown in Fig. 6a, we find that the mean restraint increases quickly with the addition of a low density of pinned particles only to show signs of saturation as c is further increased. This increase in <μ> steadily shifts our fragile alloy towards the values of restraint that characterize silica in Fig. 3. This structural shift from fragile to strong through pinning is consistent with the analogous shift in kinetic behavior reported in ref. [51]. The concentration of pinned particles required to shift <μ> by some amount can be expressed in terms of a length scale $\xi$ over which a pinned particle 'influences' its surroundings. We find the following expression,

$$< \mu(c) > = 1 - (1 - < \mu(0) >) \exp\left(-\xi c^{1/3}\right) \qquad (10)$$

provides a satisfactory description of our data and include the fitted values of the fitted length scale ξ at various values of T in Table 1. The modest increase in the length $\xi$ on cooling is similar to that reported in ref. [51]. It is useful to emphasize that this length is, by construction, purely mechanical in nature. This length does not correspond to a correlation length associated with thermal fluctuations because we do not include the pinning during the MC calculation, i.e. the pinning here is essentially just a modification of the normal analysis and so cannot effect the sampling of configurations.

|  | T=0.1 | T=0.2 | T=0.4 | T=0.6 |
|---|---|---|---|---|
| ξ (pinning) Eq. 10 | 3.58 | 3.16 | 2.62 | 2.2 |
| ζ (N dep.) Eq. 11 | 6.14 | 6.09 | 5.57 | 4.47 |
| μ(∞) Eq. 11 | 0.60 | 0.37 | 0.06 | 0.004 |

**Table 1.** The length scales ξ obtained from pinning (Eq. 10) and the length ζ and asymptotic value μ(∞) from finite size effects (Eq. 11), for the KA liquid for a number of temperatures.

A related question is the influence of system size on the amorphous solidification described by the temperature dependence of restraint. As particle restraint is a consequence of many-bodied correlations, it should exhibit some sort of characteristic length. In Fig. 6a we plot <μ> vs T for different values of N. For N < 500 we find that the T dependence of the <μ> broadens with decreasing system size and, as for pinning, the degree of restraint increases, indicting, again, a fragile-to-strong transition, this time resulting from the decrease in system size. In Fig. 6b we plot <μ> as a function of N and find the dependence of <μ> on N is described by,

$$< \mu(N) >= 1-(1-< \mu(\infty) >)\exp\left(-\zeta^3 / N\right) \tag{11}$$

where $< \mu(\infty) >$ is the asymptotic value of the mean restraint in the large N limit and the values of the associated length scale $\zeta$ are provided in Table 1. While $\zeta \sim 2\xi$, we find both length scales exhibit a similar weak increase on cooling (see Supplementary Material). Previously, Karmakar, Lerner and Procaccia [52] reported a length scale also obtained from the N dependence of the normal modes analysis of a static configuration of the KA model. Their analysis was based on the observation that the minimum eigenvalue of the inherent structure exhibits a crossover from a plastic mode to a Debye vibration as N increases, with the crossover size providing the length. The length scale $\zeta$ obtained from the N dependence of the restraint is significantly larger than that reported in ref. [52] and does not exhibit the same large increase on cooling [53]. The connection between this length and the one we extract here from the INM's is an interesting question. The crossover length of ref. [51] is found [54] to be the same as that obtained by the 'point-to-set' method [55], a variant of pinning. a result consistent with the similarity we find in the temperature dependence of the pinning and finite size scaling lengths in Table 1.

The modest increase in the length scales as provided in Table 1, along with absence of power law scaling, indicates that the structural transformation we describe in Fig. 2 is not associated with a thermodynamic singularity [56]. Since our treatment of pinning only took place in the calculation of the INM's and did not influence the sampling of the configurations, the associated length scale is, by construction, a mechanical characteristic of the instantaneous configuration. The rough correspondence between the pinning length and that of the system size effect suggests that this latter length might also be explained in mechanical terms. To clarify this point we have separated the sum for $<\Delta r_j^2>_{eq}$ in Eq.2 into the contributions from the stable and unstable modes. We can then calculate the components of restraint, $<\mu_s>$ and $<\mu_u>$, from the stable and unstable modes, respectively. Note that $<\mu> \approx <\mu_s> > <\mu_u>$ (see Supplementary Material). In Fig. 7 we plot the variation of $<\mu_s>$ and $<\mu_u>$ with respect to the pinning concentration c and the system size N.

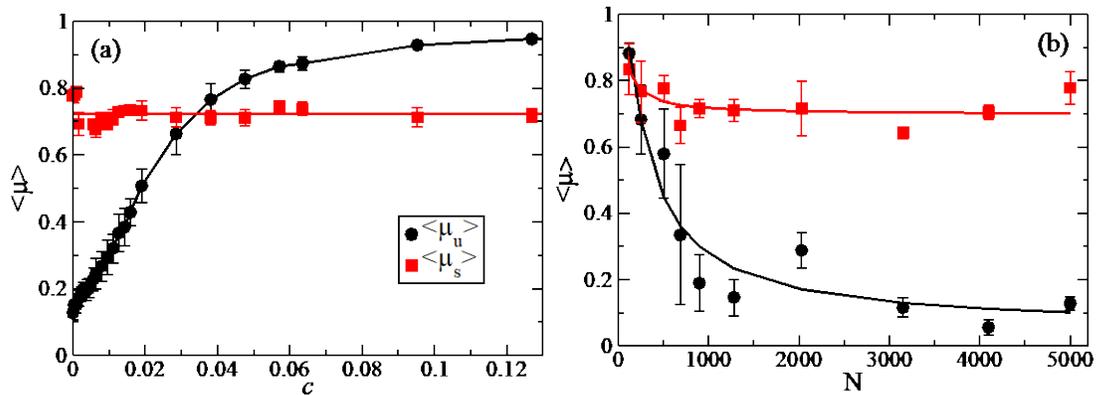

**Figure 7.** The dependence of $<\mu_s>$ and $<\mu_u>$, the average restraint contributed by stable and unstable stable modes, respectively, as a function of a) the fraction of c pinned particles and b) the system size N, for the KA liquid at T = 0.4. The data in Fig. 7a is for a N = 5000 system. The curves are provided as a guide to the eye.

The data in Fig. 7 and the Supplementary Material establishes that the increase in restraint due to the increase in the pinning concentration or the decrease in N is exerted through the suppression of the amplitude of the unstable modes. (The *number* of unstable modes, in contrast, shows only modest variation with changes in c or N, as shown in the Supplementary Material.) The stable modes are largely unaffected by the imposed constraints. In both plots in Fig. 7 we find a common crossover point at $<\mu_u> = <\mu_s> \approx 0.75$ corresponding to a value of $<\mu> (= <\mu_s> <\mu_u>) \approx 0.55$. This crossover marks the point at which the unstable modes cease to provide any reduction in restraint. These changes, achieved through variations in c or N, are, by construction, the result of direct changes to the topography of the energy landscape, as measured by the Hessian. It follows that the associated length scales in Table 1 cannot be viewed as the cause of the observed restraint but rather that restraint and its associated length scale are both characteristics of the energy landscape surface.

## 6. Conclusion and Discussion

In this paper we have demonstrated that there exists a quantifiable change in structure that provides an explicit characterization of the transformation, on cooling, of a liquid into an amorphous solid, one that can account, in purely structural terms, for the differences in vitrification of fragile and strong liquids, and the impact of imposed constraints such as pinning and finite size. Previously [44], we have established that a related measure of restraint can account for the enhanced kinetics at a glass surface.

Where the conventional account of the glass transition uses the increase of the relaxation times to measure the continuous shift of the supercooled state *away* from fluidity, the order parameter we introduce describes the solidification by measuring the continuous structural approach of the state *towards* the amorphous solid. This solid-centric approach allows us to describe amorphous solidification in a manner that is similar to the approach used to described other solidification processes. The benefits, we argue, are the technical simplification and greater versatility over descriptions based on long time dynamics, the clearer conceptual path between structure and the details of the interactions between particles and, finally, the value of aligning the description of amorphous solidification within the general theoretical framework of materials science.

Beyond the technical innovation of the restraint order parameter, we have sought to demonstrate that the associated structural phenomenology – fragility, size dependence and the pinning length scale – closely parallels the rich phenomenology previously reported in terms of the dynamics. The structural phenomenology allows for straight forward explanations. Silica is distinguished from the alloy by the large magnitude and weak temperature of the density of restraint. The length scale associated with the change in structure due to increasing the concentration of pinning particles or decreasing the size of the system is the result of the suppression of the amplitudes of the INM's. It has been proposed [57] that the transition to the amorphous solid arises as a result of the growing correlation length. Our results suggest a different interpretation, one in which the length scales are not the cause of restraint but a consequence of it.

What of dynamics? The conventional representation of amorphous solidification is based on equating solidity with a shear viscosity in excess of some threshold value, a phenomenological view that regards slow dynamics as the primary *cause* of amorphous solidification. We have argued here that solidity can be usefully treated as a structural property, implying that dynamics arises as a *consequence* of the structural transition on cooling. While there is no shortage of empirical correlations supporting this link [29-32], establishing the theoretical connection between structure and dynamics remains a hard problem, in large part because dynamics requires more information (e.g. extent of reorganization events, reversibility, etc.) than a single configuration provides. We note that the local harmonic basin we construct out of the instantaneous normal modes can provide considerably more information than that to be found in the restraint order parameter. We leave to future work the possibility of establishing a connection between structure and dynamics by exploring the nature of structural fluctuations within the harmonic Hamiltonian.

In defining an order parameter based on restraint we have a measure of considerable generality. Amorphous solidification represents a rich field of phenomena with important practical applications and unresolved fundamental questions. The nature of amorphous surfaces, the impact of atomic association ('micro alloying'), the role of molecular shape and internal flexibility, the consequences of applied strain, gelation and the precipitation of framework compounds – all of these questions can be addressed as structural problems based on the distribution of atomic restraint through straightforward applications of the methodology introduced in this paper.

**Data Availability Statement.** All the data supporting the findings of this study are available within this paper and the Supplementary Information. Additional information is available from the corresponding author upon reasonable request.

**Supplementary Material**

Included in the Supplementary Material is the following information:

1. Simulations and Normal Mode Analysis of the KA Mixture

2. Simulations of the SHK Silica and Calculation of the Hessian

3. Size dependence of Restraint in Silica

4. Effect of Pinning and System Size on the Distribution of Normal Modes

5. Calculation of Plateau Height h and its Associated Error

6. The variance of structural restraint degree on cooling of the KA mixture

7. A plot of the temperature dependence of the length scales in Table 1.

8. The accuracy of the relation $< \mu > \approx < \mu_u > < \mu_s >$

**Acknowledgements**

The authors gratefully acknowledge the assistance of Linwei Li in the generation of data for the silica system and the support from the Australian Research Council Discovery Project grant DP180104038 (PH).

**References**

1. C. S. O'Hern, L. E. Silbert, A. J. Liu and S. R. Nagel, Jamming at zero temperature and zero applied stress: The epitome of disorder. *Phys. Rev. E* **68**, 011306 (2003).

2. M. Wyart, On the rigidity of amorphous solids. *Ann. Rev. Fr.* **30**, 1-96 (2005).

3. M. van Hecke, Jamming of soft particles: geometry, mechanics, scaling and isostaticity. *J. Phys. Cond. Matter* **22**, 033101 (2010).

4. A. J. Liu and S. R. Nagel, The jamming transition and the marginally jammed solid. *Ann. Rev. Cond. Matter Phys.* **1**, 347-369 (2010).

5. A. Widmer-Cooper, H. Perry, P. Harrowell and D. R. Reichman, Irreversible reorganization in a supercooled liquid originates from localized soft modes. *Nature Phys*. **4**, 711-715 (2008).

6. J. Ding, S. Patinet, M. Falk, Y. Cheng and E. Ma, Soft spots and their structural signature in a metallic glass. *Proc. Nat. Acad. Sci USA* **111**, 14052-14056 (2014).

7. C. A. Angell, Spectroscopy simulation and scattering, and the medium range order problem in glass. *J. Non-Cryst. Sol.* **73**, 1-17 (1985).

8. T. V. Ramakrishnan and M Yussouff, First principles order-parameter theory of freezing. *Phys. Rev. B* **19**, 2775 (1979).

9. P. Tarazona, A density functional theory of melting. *Mol. Phys*. **52**, 81 (1984).

10. P. Steinhardt, D. Nelson, and M. Ronchetti, Bond orientational order in liquids and glasses. *Phys. Rev. B* **28**, 784 (1983).

11. W. Lechner and C. Dellago, Accurate determination of crystal structures based on averaged local bond order parameters. *J. Chem. Phys.* **129**, 114707 (2008).

12. Y. Hiwatari, T. Saito and A. Ueda, Structural characterization of soft-core and hard-core glasses by Delaunay tessellation. *J. Chem. Phys.* **81**, 6044 (1984).

13. J. D. Honeycutt and H. C. Andersen, Molecular dynamics study of melting and freezing of small Lennard-Jones clusters. *J. Phys. Chem*. **91,** 4950-4963 (1987).

14. S. Marin-Aguilar, H. H. Wensink, G. Foffi and F. Smallenburg, Tetrahedrality dictates dynamics in hard sphere mixtures. *Phys. Rev. Lett*. **124**, 208005 (2020).

15. K. J. Laws, D. B. Miracle and M. Ferry, A predictive structural model for bulk metallic glasses. *Nature Comm*. **6**, 8123 (2015).

16. C. P. Royall and S. R Williams, The role of local structure in dynamical arrest. *Phys. Rep.* **560**, 1-75 (2015).

17. C. De Michele, S. Gabrielli, P. Tartaglia, F. Sciortino, Dynamics in the presence of attractive patchy interactions. *J. Phys. Chem. B* **110**, 8064–8079 (2006).

18. S. Marin-Aguilar, H. H. Wensink, G. Foffi and F. Smallenburg, Slowing down supercooled liquids by manipulating their local structure. *Soft Matter* **15**, 9886-9893 (2019).

19. J. L. C. Daams and P. Villars, Atomic environment classification of the tetragonal 'intermetallic' structure types, *J. Alloys Comput.* **252**, 110 (1997).

20. D. Wei, J. Yang, M.-Q. Jiang, L.-H. Dai, Y.-J. Wang, Dyre, I. Douglass and P. Harrowell, Assessing the utility of structure in amorphous materials. *J. Chem. Phys.* **150**, 114502 (2019).

21. Y. Singh, J. P. Stoessel and P. G. Wolynes, Hard-sphere glass and the density functional theory of aperiodic crystals. *Phys. Rev. Lett.* **54**, 1059 (1985).

22. V. Lubchenko and P. G. Wolynes, Theory of structural glasses and supercooled liquids. *Ann. Rev. Phys. Chem.* **58**, 235-266 (2007).

23. G. Parisi and F. Zamponi, Mean-field theory of hard sphere glasses and jamming. *Rev. Mod. Phys.* **82**, 789-845 (2010).

24. S. Franz and G. Parisi, Recipes for metastable states in spin glasses. *J. Phys. I* **5**, 1401-1415 (1995)

25. B. Guiselin, G. Tarjus and L. Berthier, On the overlap between configurations in glassy liquids. *J. Chem. Phys.* **153**, 224502 (2020).

26. B. Guiselin, G. Tarjus and L. Berthier, Static self-induced heterogeneity in glass-forming liquids: Overlap as a microscope. arXiv:2201:10183 (2022).

27. R. M. Stratt, The instantaneous normal modes of liquids. *Acc. Chem. Res.* **28**, 201-207 (1995).

28. B. Maden and T. Keyes, Unstable modes in liquids density of states, potential energy, and heat capacity. *J. Chem. Phys*. **98**, 3342 (1993).

29. J. C. Dyre, T. Christensen and N. B. Olsen, Elastic models for the non-Arrhenius viscosity of glass-forming liquids. *J. Non-Cryst. Sol*. **352**, 4635-4642 (2006).

30. L. Larini, A. Ottochian, C. De Michele and D. Leporini, Universal scaling between structural relaxation and vibrational dynamics in glass-forming liquid and polymers. *Nature Phys.* **4**, 42-45 (2008).

31. D. S. Simmons, M. T. Cicerone, Q. Zhong, M. Tyagi and J. F. Douglas, Generalized localization model of relaxation in glass-forming liquids. *Soft Matter* **8**, 11455(2012).

32. U. Buchenau, R. Zorn and M A. Ramos, Probing cooperative liquid dynamics with the mean squared displacement. *Phys. Rev. E* **90**, 042312 (2014).

33. X. Yang, H. Tong, W.-H. Wang and K. Chen, Emergence and percolation of rigid domains during the colloid glass transition. *Phys. Rev. E* **99**, 062610 (2019).

34. J. Ding, Y.-Q. Cheng, H. Sheng, M. Asta, R. O. Ritchie and E. Ma, Universal structural parameter to quantitatively predict metallic glass properties. *Nature Com*. **7**, 13733 (2016).

35. S. S. Schoenholz, E. D. Cubuk, E. Kaxiras and A. J. Liu, Relationship between local structure and relaxation in out-of-equilibrium glassy systems. *Proc. Nat. Acad. Sci*. **114**, 263 (2017).

36. H. Tong and N. Xu, Order parameter for structural heterogeneity in disordered solids. *Phys. Rev. E* **90**, 010401(R) (2014).

37. A. Smessaert and J. Rottler, Structural relaxation in glassy polymers predicted by soft modes: a quantitative analysis. *Soft Matter* **10**, 8533 (2014).

38. W. Kob and H. C. Andersen, Testing mode-coupling theory for a supercooled binary Lennard-Jones mixture – The van Hove correlation function. *Phys. Rev.* **51**, 4626-4641 (1995).

39. T. S. Grigera and G. Parisi, Fast Monte Carlo algorithm for supercooled soft spheres. *Phys. Rev. E* **63**, 045102 (2001).

40. C. Donati, F Sciortino and P. Tartaglia, Role of unstable directions in the equilibrium and ageing dynamics of supercooled liquids. *Phys. Rev. Lett*. **85**, 1464 (2000).

41. G. Sun, L. Li and P. Harrowell, The structural difference between a strong and a fragile liquid. *J. Non-Cryst. Solids* **13**, 100080 (2022).

42. U. R. Pedersen, T. B. Schröder and J. C. Dyre, Phase diagram of Kob-Andersen-type binary Lennard-Jones mixtures. *Phys. Rev. Lett*. **120**, 165501 (2018).

43. C. Donati, J. F. Douglass, W. Kob, S. J. Plimpton, P. H. Poole and S. C. Glotzer, Stringlike cooperative motion in a supercooled liquids. *Phys. Rev. Lett*. **80**, 2338 (1998).

44. G. Sun, S. Saw, I. Douglass and P. Harrowell, Structural origin of enhanced dynamics at the surface of a glassy alloy. *Phys. Rev. Lett*. **119**, 245501 (2017).

45. W. Gotze and L. Sjogren, Relaxation processes in supercooled liquids. *Rep. Prog. Phys*. **55**, 241 (1992).

46. S. Sundararaman, L. Huang, S. Ispas, W. Kob, New optimization scheme to obtain interaction potentials for oxide glasses. *J. Chem. Phys.* **148**, 194504 (2018).

47. B. A. Luty, M. E. Davis, I. G. Tironi and W. F. van Gunsteren, A comparison of particle-particle, particle-mesh and Ewald methods for calculating electrostatic interactions in periodic molecular systems. *Mol. Sim.* **14**, 11 (1994).

48. A. Takada, P. Richet, C. R. A. Catlow and G. D. Price, Molecular dynamics simulations of vitreous silica structures. *J. Non-Cryst. Solids* **345-346**, 224-229 (2004).

49. C.A.Angell, Glass-formers and viscous liquid slowdown since David Turnbull: Enduring puzzles and new twists. *MRS Bulletin*, **33**, 544-555 (2008).

50. L. Angelani, M. Paoluzzi, G. Parisi and G. Ruocco, Probing the non-Debye low-frequency excitations in glass through random pinning. *Proc. Nat. Acad. Sci.* **115**, 8700-8704 (2018).

51. S. Chakrabarty, S. Karmakar and C. Dasgupta, Dynamics of glass forming liquids with randomly pinned particles. *Sci. Rep.* **5**, 12577 (2015).

52. S. Karmakar, E. Lerner and I. Procaccia, Direct estimate of the static length-scale accompanying the glass transition. *Physica A* **391**, 1001-1008 (2012).

53. R. Gutiérrez, S. Karmakar, Y. G. Pollack and I. Procaccia, The static lengthscale characterizing the glass transition at lower temperatures. *EPL* **111**, 56009 (2015).

54. G. Biroli, A. Karmakar and I. Procaccia, Comparison of static length scales characterizing the glass transition. *Phys. Rev. Lett.* **111**, 165701 (2013).

55. J.-P. Bouchaud and G. Biroli, On the Adam-Gibbs-Kirkpatrick-Thirumalai-Wolynes scenario for the viscosity increase in glasses. *J. Chem. Phys*. **121**, 7347 (2004).

56. K.Binder, Finite size effects on phase transitions. *Ferroelectrics* **73**, 43-67 (1987).

57. S. Karmakar, C. Dasgupta and S. Sastry, Growing length scales and their relation to timescales in glass-forming liquids. *Ann. Rev. Cond. Matter Phys*. **5**, 255-284 (1014).

# Supplementary Material

# A General Structural Order Parameter for the Amorphous Solidification of a Supercooled Liquid

Gang Sun, Peter Harrowell*

* Correspondence to: peter.harrowell@sydney.edu.au

**This file includes:**

1. Simulations and Normal Mode Analysis of the KA Mixture

2. Simulations of the SHK Silica and Calculation of the Hessian

3. Size dependence of Restraint in Silica

4. Effect of Pinning and System Size on the Distribution of Normal Modes

5. Calculation of Plateau Height h and its Associated Error

6. The variance of structural restraint degree on cooling of the KA mixture

7. A plot of the temperature dependence of the length scales in Table 1.

8. The accuracy of the relation $< \mu > \approx < \mu_u > < \mu_s >$

## 1. Simulations and Normal Mode Analysis of the KA Mixture

The Monte Carlo simulations of the $A_{80}B_{20}$ Lennard-Jones mixture use the potential introduced by Kob and Andersen [Phys. Rev. 51, 4626 (1995)] with 5000 particles in a simulation cell with periodic boundary conditions. Well relaxed configurations are generated using particle swap Monte Carlo (MC) simulations at constant NVT with a fixed number density of 1.2.

To simulate the systems, we have used a Monte Carlo algorithm from the DLMONTE 2.0 version suite of software. The random walk in the MC simulations is constructed using a Metropolis scheme that iterates the following three steps:

1. Select a particle $i$ at random and calculate its energy $U(\mathbf{r}_1)$.

2. Displace the particle randomly to a *trial* position and calculate its new energy $U(\mathbf{r}_2)$.

3. Accept or reject the particle displacement according to the Metropolis acceptance rule:

$P$ acc($\mathbf{r}_1 \rightarrow \mathbf{r}_2$)=min(1,exp$\{-\beta[U(\mathbf{r}_2)-U(\mathbf{r}_1)]\}$)

Swap moves are also applied in this work, with the exchange frequency n=50 time steps. The details are as follows:

4. Choose a random particle $j$ of a different type.

5. Swap particle positions $\mathbf{r}_j \leftrightarrow \mathbf{r}_i$ . Shift both particles as in the standard move

6. Accept or reject new configuration according to the Metropolis criterion

7. Repeat for all particles.

The calculation of the variance of the amplitude $A_k$ of mode k for cases where the eigenvalue $\lambda_k < 0$ were carried out as follows.

A 1D energy curve can be plotted based on the variation of the potential energy U with respect to the selected amplitude $A_k$ while setting all other mode amplitudes to zero. The results for two such curves are shown in Fig. 1b, one with $\lambda_k < 0$ and one with $\lambda_k > 0$. In the case of the $\lambda_k < 0$ mode we see the region of negative curvature. While the value of $< A_k^2 >$ can be calculated analytically for the $\lambda_k > 0$ case, as indicated by Eq. 3 in the paper, $< A_k^2 >$ must be calculated numerically when $\lambda_k < 0$. As can be seen from Fig. 1b, this variance is well defined due to the positive contributions of higher order terms in $A_k$. The numerical integration indicated in Eq. 4 is carried out over the range of $A_k$ for which $\Delta E \leq 10k_B T$. As shown in Fig. S1, increasing this threshold produces no change in the calculated value.

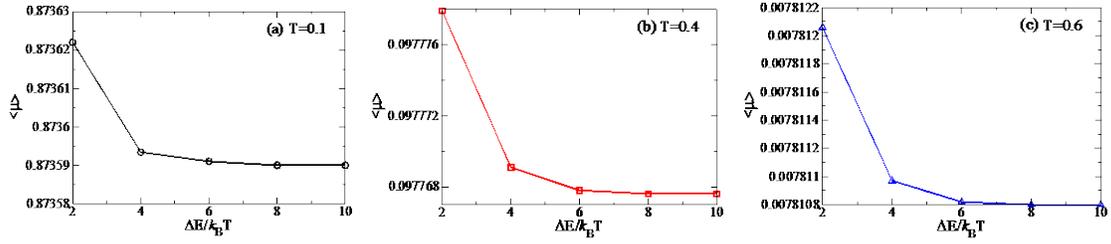

**Figure S1**. The dependence of the mean restraint $\langle\mu\rangle$ on the energy cut-off $\Delta E/k_B T$ at different temperatures. In these calculations, the amplitudes of all modes are evaluated by numerical integration of Eq. 4. We find no change in $\langle\mu\rangle$ when $\Delta E/k_B T$ is increased beyond a value of 10, the cut-off used in the calculations presented in this paper.

The capacity of the MC algorithm to equilibrate a system's is conditional on the temperature and the associated relaxation rate. To establish the lowest temperature at which we are generating fully equilibrated configurations, we have calculated the average potential energy $E_p$ as a function of temperature for different cooling rates q, defined as the change in the reduced temperature per C cycle. The results are plotted in Fig. S2. We find that at the quench rate, $q = 4 \times 10^{-7}$, used in the calculations in this paper, equilibration is achieved down to $T = 0.36$. Below this temperature we see the deviation in energy shown in Fig. S2.

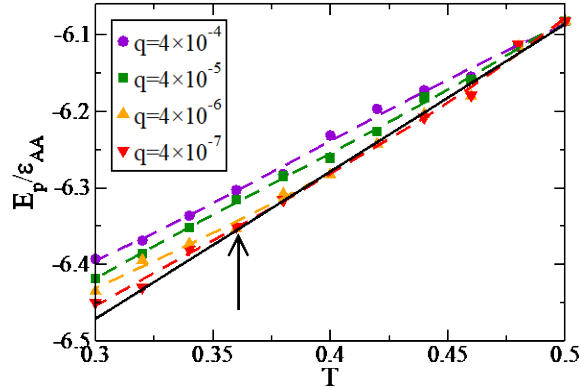

**Figure S2.** The temperature dependence of the potential energy $E_p$ of the KA mixture relaxed at various cooling rates q, as indicated. The arrow indicates the temperature (=0.36) where the average potential energy deviates from equilibrium at the cooling rate $q = 4 \times 10^{-7}$.

In Fig. S3 we plot the temperature dependence of the fraction of modes with negative eigenvalues and the distributions of average square $\langle A_\alpha^2\rangle$'s from the stable and unstable modes. We find that the fraction of unstable modes decreases monotonically with decreasing T and vanishes at $T \sim 0$. As shown in Fig. S2, the amplitudes associated with the unstable modes, evaluated using Eq. 4, are very much larger than those of the stable modes and so contribute significantly to the magnitude of $< \Delta r_j^2 >$ .

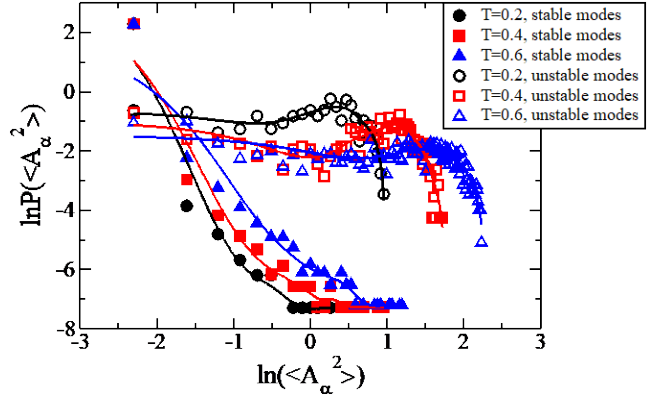

**Figure S3.** The log-log plot of the distribution of mode amplitudes $< A_\alpha^2 >$'s from the stable and unstable modes at T = 0.6, 0.4 and 0.2. The curves are fits included as a visual guide.

## 2. Simulations of the SHK Silica and Calculation of the Hessian

We perform the molecular dynamics (MD) simulations of silica by using SHIK-1 potentials [J. Chem. Phys. 148, 194504 (2018)] which have high geometric accuracy. This potential includes long range Coulomb interactions requiring an Ewald sum. The silica systems contain 576 atoms. All the samples are prepared using the melt-quench method. At first, we generate an Alpha-quartz model at 300 K, then we relaxed the quartz at 300K, zero pressure condition for 0.1 ns to release the pressure. After the quartz was relaxed, we heated the quartz to 4000 K at a constant rate over 0.5 ns, then equilibrated the melted sample at a density of 2.2 g/cm³ for another 0.5 ns in the NVT ensemble. The sample was then quenched to 300 K in the NPT ensemble at a rate of 1 K/ps with zero pressure. After quenching, it was then annealed at 300 K for another 1 ns.

Hessian matrix elements are usually defined as,

$$H_{i\mu,j\upsilon} = \frac{1}{\sqrt{m_i m_j}} \frac{\partial^2 U}{\partial r_{i,\alpha} \partial r_{j,\beta}} (i,j = 1,...,N, \mu,\upsilon = x,y,z) \qquad (S1)$$

Where U is the total potential energy of the system. Since long range interactions are involved in the simulation of silica, analytical expressions for the Hessian matrix elements are complicated. Therefore, we generated the Hessian matrix moving the individual particles from their starting configuration $R$ to $R+\delta r_{j,\upsilon}$ in order to counterbalance, a double move of $R$ to $R$ $-\delta r_{j,\upsilon}$, is employed. The displacement of 0.0016 Å has been chosen as it is small enough to avoid any disruption to the structure and gives reasonably accurate result.

$$\sqrt{m_i m_j} H_{i\mu,j\upsilon} = \nabla_{i,\mu}\nabla_{j,\upsilon}U = \frac{\psi_{i,\mu}(R+\delta r_{j,\upsilon}) - \psi_{i,\mu}(R-\delta r_{j,\upsilon})}{2\delta r_{j,\upsilon}} \qquad (S2)$$

where U is the potential energy and $\psi_{i,\mu}$ is the force of particle $i$ along $\mu$ direction.

On diagonalizing the Hessian we find a fraction f of modes with negative eigenvalues that decreases monotonically with temperature and appears to vanish at T = 0 as shown in Fig. S4.

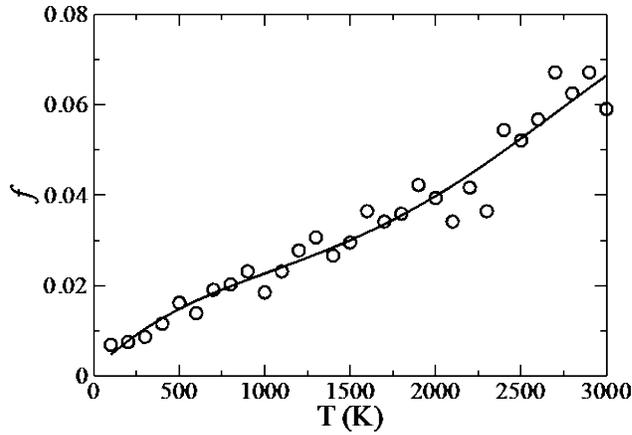

**Figure S4**. The fraction of unstable modes of silica as function of T.

### 3. Size Dependence of Restraint in Silica

Calculations of restraint for silica for a range of systems sizes N are planned for future work. In the meantime, we argue that the value of N used in this paper provides us with a reasonable estimate of the asymptotic behavior of restraint in the $SiO_2$ system. The argument is based on the similar trends observed between the correlation lengths from system size and pinning calculations in the KA model. According to the fitted function, Eq.11, the asymptotic size dependence for the KA systems is achieved when N $\gg \zeta^3$. From Table 1 we can assign $\zeta$ its maximum value of 6.14 so that for the N = 5000 system, N/$\zeta^3$ = 21, a value large enough safely set $<\mu(N)>\sim<\mu(\infty)>$. We can calculate an analogous length $\xi$ (see Eq. 10) from pinning with $\zeta \sim 2\xi$. Using this connection, the condition that we have achieved the asymptotic behavior in restraint can be written using the pinning length, i.e. $N/(2\xi)^3 \gg 1$. We have calculated the pinning correlation length for silica in ref. 41. The values of $\xi$ for $SiO_2$ are presented in Table S1 in units of the Si-O bond length (1.59 Å). In the current paper we have used a system size N = 516 for silica. The ratio $N/(2\xi)^3$ for N = 516 is also presented in

Table S1. We find that the asymptotic condition $N/(2\xi)^3 >> 1$ is met for $T/T_m \geq 0.13$. As the substantial difference we report between restraint in $SiO_2$ and the KA mixture involves the high temperatures, we argue that the selected system size is sufficient for capturing the large N limit.

| $T/T_m$ | 0.04 | 0.13 | 0.38 | 0.63 | 0.88 | 1.25 |
|---|---|---|---|---|---|---|
| $\xi$ | 3.14 | 1.32 | 0.94 | 0.63 | 0.44 | 0.25 |
| $N/(2\xi)^3$ | 2 | 28 | 78 | 257 | 757 | $>10^3$ |

**Table S1.** The pinning length $\xi$ (in units of the Si-O bond length of 1.59 Å) for different values of T relative to the melting point $T_m$=2400K.

4. Effect of Pinning and System Size on the Distribution of Normal Modes

We randomly select the particles for pinning, taking no steps to avoid Poisson clustering of pinning sites. We find that the pinning concentration and system size exhibits a modest impact in the density of states of the INM's. In Fig. S5 a and b we plot the density of states of the INM's at T = 0.4 for different values of c and N, respectively. In Fig. 5 c and d we plot the fraction f of modes with an eigenvalue $\lambda_\alpha < 0$ as functions of c and N. While we find the fraction f exhibits a modest increase with increasing c, an opposite trend is observed when decreasing N. This increase in f with decreasing system size is similar to that reported by Lerner [E. Lerner, Phys. Rev. E 101, 032120 (2020)]

As discussed in the text, the main impact of the imposed constraints is felt through the amplitude of the INM's and the resultant magnitudes of $<\Delta r_i^2>_{eq}$ and $\mu_i$. The trends in the variance of the mode amplitudes $<A_\alpha^2>_{eq}$ are plotted in Fig. S6 for the stable and unstable modes as a function of c and N.

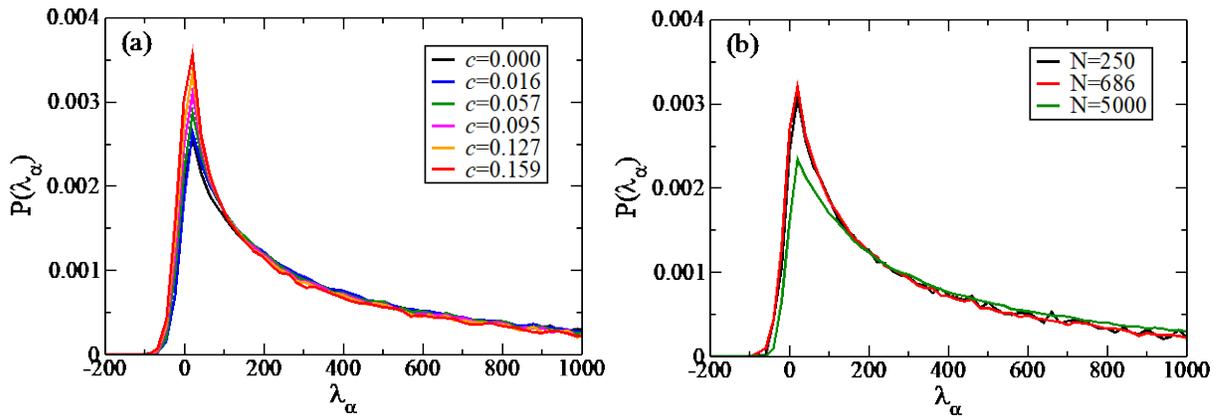

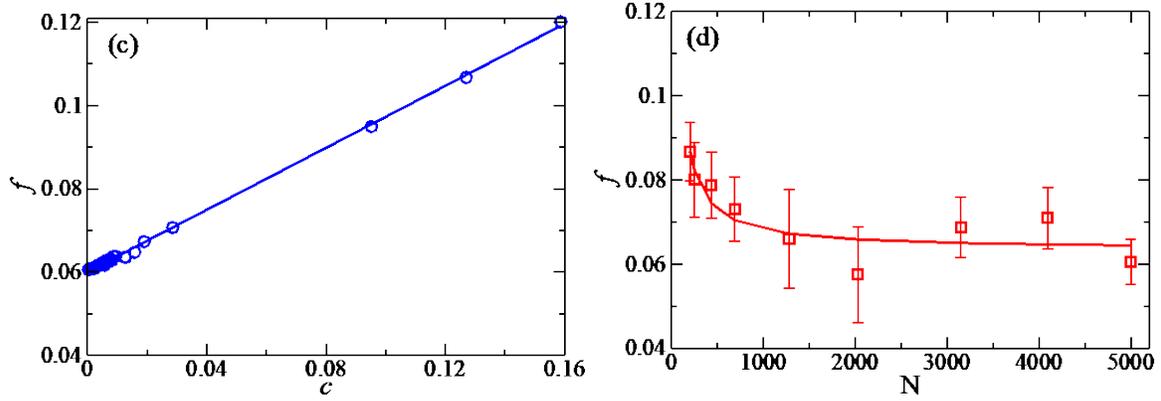

**Figure S5.** The distribution of eigenvalues of the INM's of the KA model at T = 0.4. Distributions for different values of a) the pinning concentration c (for N = 5000), and b) the system size N, are shown. The dependence of the fraction of unstable modes f on the values of c) the pinning concentration and d) the system size are also shown.

The above results establish that the addition of pinned particles does not substantially change the density of states or, more specifically, the fraction of unstable modes, and that the decrease in system size sees an *increase* in unstable modes, in contrast to the observed increase in restraint shown in Fig. 6b. To understand how these variables do exert their influence on restraint, we nest consider the distribution of the means squared mode amplitudes $<A_\alpha^2>_{eq}$ calculated as described in the main text. The results for the distribution of $<A_\alpha^2>_{eq}$ for the stable and unstable modes are shown In Fig. S6 for a number of values of c and N.

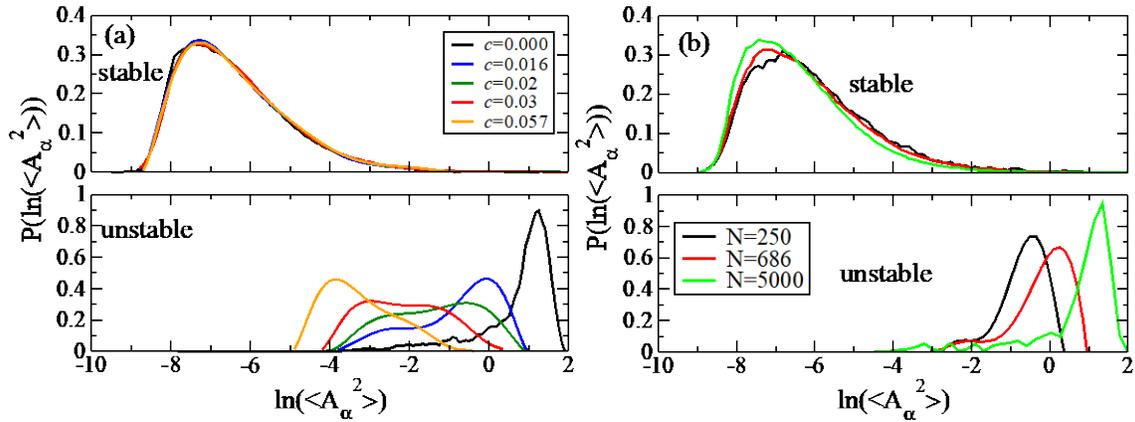

**Figure S6.** The distribution of mode amplitudes $<A_\alpha^2>$'s from the stable and unstable modes for systems with pinning particles c=0, 0.016, 0.02, 0.03 and 0.057 (for N = 5000) (a), and size N=250, 686, 5000 (b), at T = 0.4.

## 5. Calculation of Plateau Height h and its Associated Error

For supercooled liquids, the self intermediate scattering function Fs(q,t), is taken at the a value of the wavevector $q=q_1$, which corresponds to the first peak of the structure factor S(q). The structure factor as function of wavevector is shown in Fig.S7.

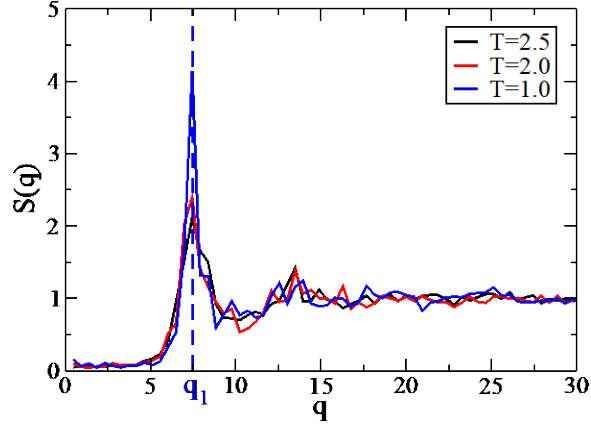

**Figure S7.** The structure factor S(q) as function of wavevector q for T=2.5, 2.0 and 1.0. The first peak position q is 7.45.

For liquids, the relaxation occurs in two steps, observed from the self intermediate scattering function $F_s(q_1,t)$ [see Fig.S8].The two step decays correspond to α relaxation (slow) and β relaxation (fast). Hence, it is reasonable to fit $F_s(q,t)$ of liquids by a sum of two functions,

$$F_s(q_1,t) = (1-h)\exp\left[-\left(\frac{t}{\tau_\beta}\right)^2\right] + h\exp\left[-\left(\frac{t}{\tau_\alpha}\right)^2\right]$$

(S3)

where $h$ is the plateau height of relaxation function $F_s$. The quality of the fits is demonstrated in two examples plotted in Fig. S8.

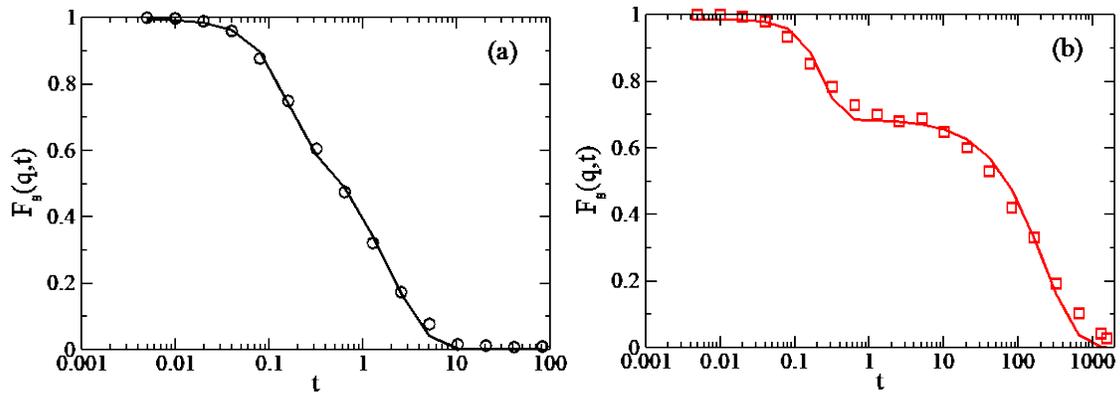

**Figure S8.** Self intermediate function $F_s(q,t)$ for T=1.0 (a) and 0.5 (b). The solid curves are the ones fitted by Eq.S3.

The quality of the fit is measured by the correlation coefficient ε between the fitted curves and the intermediate scattering function $F_s(q_1,t)$. What we would like to establish is how ε varies with the value of h. To estimate the uncertainty in the value of we have proceeded as follows.

Step 1. Fit the intermediate scattering function $F_s(q_1,t)$ with the Eq.(S3), and let h=h*.

Step 2. Fix h=h*+δ and redo the fit over the 2 remaining variables and record the overall measure of the quality ε of the new fit.

Step3. Repeat this for a range of δ's, +ve and -ve, and plot ε as a function of δ. The result is plotted in Fig. S9. As expected, there is an extremum at δ = 0, i.e. the original fit returned the best value of h. What we are interested in is the curvature about the extremum. If it is high curvature then it means that the uncertainty in the original value of h was small. If the curvature is low, i.e. varying h does not change the quality of fit, then the uncertainty in h is high. The uncertainty in *h* can be estimated by the interval in δ over which ε lies above some threshold value, here $\varepsilon^* > 0.95$.

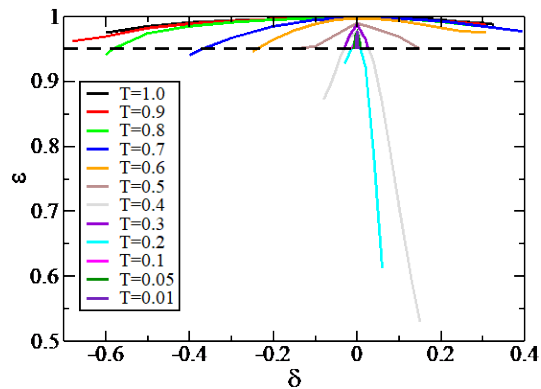

**Figure S9.** The quality of the fits as a function of δ (explained in the text) for all temperatures in this work. The black dashed line is the threshold value chosen here, ε*=0.95.

6. The variance of structural restraint degree on cooling of the KA mixture

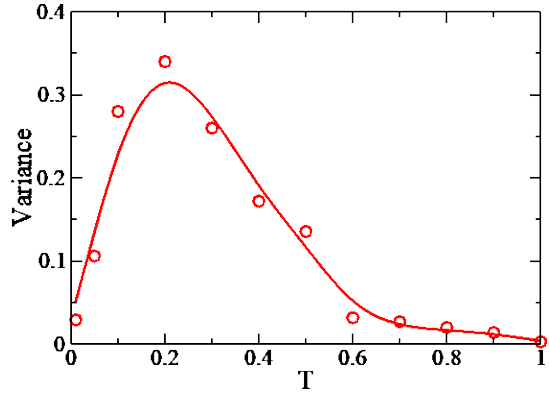

**Figure S10.** The variance of structural restraint degree (μ) of KA model as the function of temperature on cooling process.

7. <u>A plot of the temperature dependence of the length scales in Table 1</u>

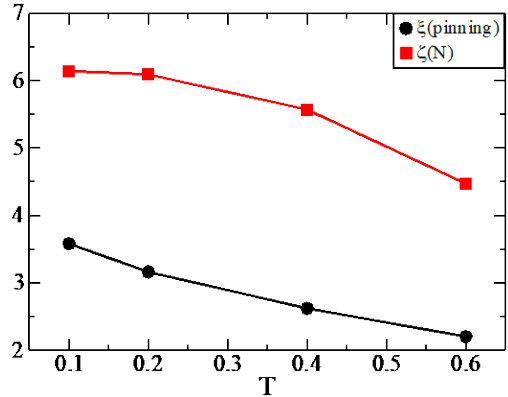

**Figure S11.** The length scales obtained from pinning and size dependence as listed in Table 1 as function of temperature.

8. <u>The accuracy of the relation</u> $< \mu > \approx < \mu_u > < \mu_s >$

In the text of Section 6 we resolve the average restraint into contributions from the stable $< \mu_s >$ and unstable $< \mu_u >$ modes. It is stated in the text that these individual component averages are related to the overall restraint by $< \mu > \approx < \mu_u > < \mu_s >$. In Fig. S12 we confirm this relation numerically for the results from the pinning and system size dependent calculations.

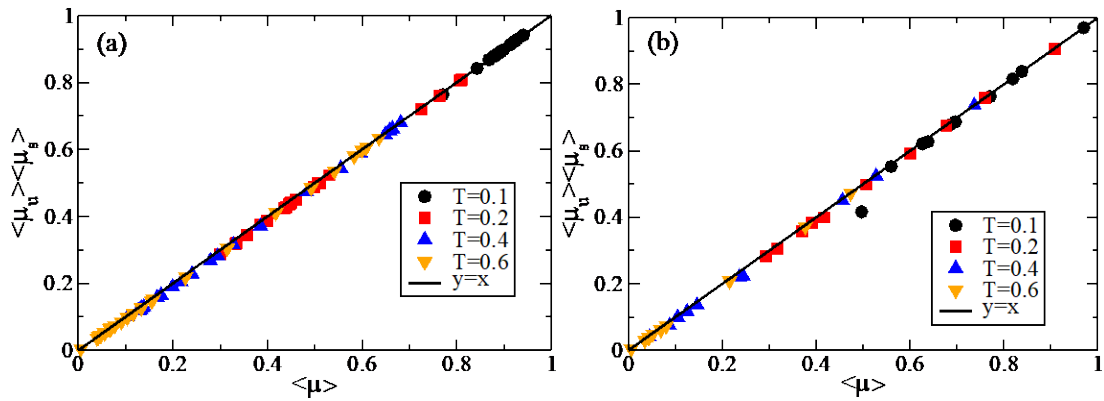

**Figure S12.** $<\mu_u><\mu_s>$ plotted against the average overall restraint $<\mu>$ for systems with different pinning concentrations (a) and different system size N (b) for a range of temperatures.